\documentclass[epj,final]{svjour}
\usepackage{epsfig}

\begin{document}
\onecolumn
\title{Search for Resonant Absorption of Solar Axions Emitted in M1 Transition in $^{57}$Fe Nuclei.}

\author{
A.V.~Derbin,{\thanks{derbin@pnpi.spb.ru}} A.I.~Egorov, I.A.~Mitropol'sky, V.N.~Muratova, D.A.~Semenov, E.V.~Unzhakov}

\date{}
\twocolumn
 \institute { {{St.Petersburg Nuclear Physics Institute, 188300 Gatchina, Russia}}\\
 }

\abstract{ A search for resonant absorbtion of 14.4 keV solar axions by $^{57}$Fe target was performed. The Si(Li) detector
placed inside the low-background setup was used  to detect the $\gamma$-quanta appearing in the deexcitation of 14.4 keV nuclear
level: $A+\rm{^{57}Fe} \rightarrow \rm{^{57}Fe^{*}} \rightarrow \rm{^{57}Fe} + \gamma$. The new upper limit for the hadronic
axion mass have been obtained: $m_{A} \leq 151$ eV (90\% C.L.) ($S$=0.5, $z$=0.56). {}}

\titlerunning{Search for the resonant absorption of solar axions...}
\authorrunning{A.V.~Derbin, A.I.~Egorov et al.}
\maketitle

\section{Introduction}
A natural solution of the strong CP problem based on the global chiral symmetry U(1) was proposed by Peccei and Quinn
\cite{Pec77}. Weinberg \cite{Wei78} and Wilczek \cite{Wil78} noted that spontaneous breaking of the PQ-symmetry at the energy
$f_A$ leads to the existence of a new neutral spin-zero pseudoscalar particle - axion. The axion mass ($m_A$) and the strength
of an axion coupling to an electron ($g_{Ae}$), a photon ($g_{A\gamma}$) and nucleons ($g_{AN}$) are proportional to the inverse
of $f_A$. The original PQWW-model with $f_A$ fixed at the electroweak scale $f_A$ = $(2G_F)^{-1/2}$, was excluded after
intensive experimental searches which were performed using radioactive sources, reactors and accelerators (see \cite{PDG08} and
refs therein).

New axion models decoupled the scale of PQ symmetry breaking from the electroweak scale and the value of $f_{A}$ can be extended
up to the Plank mass $\approx$10$^{19}$ GeV. Two classes of models for the "invisible" axion have been developed: the KSVZ or
hadronic axion model \cite{Kim79},\cite{Shi80} and the DFSZ or GUT axion models \cite{Zhi80},\cite{Din81}.

The axion mass in both models is given in terms of neutral pion properties:
\begin{equation}\label{ma}
  m_A=\frac{f_\pi m_\pi}{f_A}(\frac{z}{(1+z+w)(1+z)})^{1/2}
\end{equation}
where $f_\pi$ $\cong$ 93 MeV is the pion decay constant, $z =m_u/m_d \approx 0.56$ and $w = m_u/m_s \approx 0.029$ are
quark-mass ratios. For the given values of $z$ and $w$ the equation (\ref{ma}) is numerically presented as $m_A\cong
6.0\cdot10^{6} / f_A $ where $m_A$ and $f_A$ are in eV and GeV units, correspondingly.

In contrast to the DFSZ axions, the KSVZ axions have no coupling to leptons and ordinary quarks at the tree level, which results
in the strong suppression of the interaction of the KSVZ axion with electrons through radiatively induced coupling \cite{Sre85}.
Moreover, in some models axion photon coupling may differ from the original DFSZ or KSVZ $g_{A\gamma}$ couplings by a factor
less than 10$^{-2}$ \cite{Kap85}.

The axion coupling constants are constrained by various experiments, astrophysical and cosmological arguments. The astrophysical limits
based on the axion interaction strength with photons and electrons in stars suggest that $m_A <$ 0.01 eV if one assumes the standard
$g_{A\gamma,Ae}-m_A$ relations (see \cite{PDG08,Raf06} and refs therein).

The axion-nucleon coupling is constrained by the upper and lower limits based on the observed neutrino signal of SN1987A
\cite{Raf99}. The supernova data leave open the so called "hadronic axion window" of $m_A \sim$  10 eV if $g_{A\gamma}$ coupling
is sufficiently small. However, the cosmological limits on hot dark matter consisting of axions provide $m_A<$1 eV \cite{Raf08}.

If axions do exist, the Sun would be an intense source of these particles. Several main mechanisms of solar axion production are
considered. Axions can be efficiently produced by Primakoff conversion of photons in the electric field of plasma. The resulting
axion flux has an average energy of about 4 keV and can be detected by inverse Primakoff conversion in laboratory magnetic
fields \cite{Duf06} - \cite{Ari09} or by the coherent conversion to photons in crystal detectors \cite{Avi99}-\cite{Ahm09}. The
experiments are sensitive to $g_{A\gamma}$ coupling. The obtained upper limits for the axion-photon coupling constant are
$g_{A\gamma} \leq10^{-10}\div 10^{-8}$ GeV$^{-1}$, which still corresponds to the immense estimated axion flux at the level of
10$^{11}\div$10$^{13}$ cm$^{-2}$s$^{-1}$keV$^{-1}$.

The other source of axions are reactions of the solar cycle that produce solar energy \cite{Raf82}. Since axions are
pseudoscalar particles they can be emitted in nuclear magnetic transitions. Attempts to detect 478 keV monochromatic axions
through the resonant absorption in $^{7}$Li nuclei target have been performed in \cite{Krc01}-\cite{DAMA08}.

Since the temperature in the center of the Sun is $\sim1.3$ keV and some nuclei having low- lying nuclear levels can be excited
thermally \cite{Hax91}. Monochromatic axions can be emitted in the magnetic nuclear transitions from the first thermally excited
level to the ground state. The aim of the present work is to search for the 14.4 keV solar axions emitted by the M1-transition
in $^{57}$Fe nuclei. The axions on the Earth can be observed in the inverse reaction of the resonant absorption via the
registration of $\gamma$-rays (or conversion electrons) produced by the discharge of the excited nuclear level \cite{Mor95}. The
probability of emission and subsequent absorption of the axion in a magnetic transition is determined only by the axion-nucleon
coupling. The previous searches for 14.4 keV axion were performed in \cite{Krc98}-\cite{Der07_A}.

\section{Emission and absorption of axions in nuclear transitions of magnetic type.}
As it has been found in \cite{Hax91}, the most intense solar axion flux is connected with M1-transition of $^{57}$Fe nucleus.
The energy of the first excited nuclear level $3/2^{-}$ is equal to 14.413 keV, and the admixture of the E2 transition is
$\delta^2 = 0.22\%$. The total axion flux $\Phi_{A}$ depends on the level energy $E_{\gamma}$= 14.413 keV, temperature $T$,
nuclear level lifetime $\tau_{\gamma}$ = 1.34 $\mu$s, the abundance of the $^{57}$Fe isotope on the Sun $N$ and the branching
ratio of axions to photons emission $\omega_A /\omega_\gamma$ \cite{Hax91,Mor95}:

\begin{equation}\label{eq2}
\Phi_A \sim\frac{N}{\tau_\gamma }\frac{2 exp(-E_\gamma /kT)}{(1+2 exp(-E_\gamma /kT))}\frac{\omega_A}{\omega_\gamma}
\end{equation}

Owing to the Doppler broadening, the axion spectrum is a sum of Gaussian curves $\Phi_A(E_A)$ with the dispersion $\sigma(T)
=E_\gamma(kT/M)^{1/2}$, where T is the temperature at the point where the axion is emitted and M is the mass of the $^{57}$Fe
nucleus. The axion flux was calculated for the temperature dependence on the radius given by BS05(OP) Standard Solar Model
\cite{Bah05} based on the corona high-Z abundances \cite{Gre98}. At the Earth, the differential axion flux at the maximum of the
quasi-Gaussian distribution can be presented as:

\begin{equation}\label{axionflux_num}
\Phi_{A}(E_{M1}) = 4.15\times 10^{25}\left(\frac{\omega_{A}}{\omega_{\gamma}}\right) \mbox{cm}^{-2} \mbox{s}^{-1}
\mbox{keV}^{-1}.
\end{equation}
The obtained width of the axion line is equal to $\sigma_{S} = 2.2$ eV. This value significantly exceeds the energy of
recoil-nucleus (1.8 $\mu$eV), as well as the Doppler broadening of the line at temperature T = 300 K of the target nuclei (10
meV), and the self width of the level $\Gamma = 4.65 \cdot 10^{-9}$ eV. Thus, the percentage of axions satisfying the
resonant-absorption condition amounts to the value $\sim\Gamma/\sigma_{S}$.

Within the framework of the long-wavelength approximation, the axion emission probability $(\omega_{A}/\omega_{\gamma})$ is given by the
expression \cite{Don78,Avi88,Hax91}:
\begin{equation}\label{axion_prob}
\frac{\omega_{A}}{\omega_{\gamma}} =
\frac{1}{2\pi\alpha}\frac{1}{1+\delta^2}\left[\frac{g^{0}_{AN}\beta+g^{3}_{AN}}{(\mu_{0}-0.5)\beta+\mu_{3}-\eta}\right]^{2}
\left(\frac{p_{A}}{p_{\gamma}}\right)^{3},
\end{equation}
where $p_{\gamma}$ and $p_{A}$ are the photon and axion momenta respectively; $\alpha\approx1/137$, $\mu_{0}\approx0.88$,
$\mu_{3}\approx4.71$ are the isoscalar and isovector nuclear magnetic momenta, $\beta$ and $\eta$ are the parameters depending on the
nuclear matrix elements. The values $\beta=-1.19$ and $\eta=0.8$ for the M1 transition in the $^{57}$Fe nucleus were calculated in
\cite{Hax91}.

The interaction of the axion with nucleons is determined by the coupling constant $g_{AN}$, which consists of isoscalar
$g^{0}_{AN}$ and isovector $g^{3}_{AN}$ parts. In the hadronic axion models, $g^{0}_{AN}$ and $g^{3}_{AN}$ constants can be
represented in the form \cite{Sre85,Kap85}:
\begin{equation}\label{couplcons}
g_{AN}^{0}=-\frac{m_N}{6f_A}[2S+(3F-D)\frac{1+z-2w}{1+z+w}],
\end{equation}
\begin{equation}
g_{AN}^{3}=-\frac{m_N}{2f_A}[(D+F)\frac{1-z}{1+z+w}].
\end{equation}
Here, $m_{N}$ = 939 MeV is the nucleon mass, the constants $D$ and $F$ are expressed in terms of the isovector ($F_{A3}$) and isoscalar
($F_{A0}$) pion-nucleon coupling constants. The exact values of $D$ and $F$ parameters determined from the semileptonic hyperon decays are
equal to $D = 0.808\pm0.006$ and $F = 0.462\pm0.011$ \cite{Mat05}.

The parameter $S$ characterizing the flavor singlet coupling still remains a poorly constrained one. Its value varies from
$S=0.68$ in the naive quark model down to $S=-0.09$ which is given on the basis of the EMC collaboration measurements
\cite{May89}. The more stringent boundaries $(0.37\leq S\leq0.53)$ and $(0.15\leq S\leq0.5)$ were found in \cite{Alt97} and
\cite{Ada97}, accordingly. As a result the value of the sum ($g_{AN}^0\beta+g_{AN}^3$) in (\ref{axion_prob}) may significantly
decreases and, due to negativity of the parameter $\beta$, actually vanishes. Taking into account that the usually accepted
value of $u$- and $d$-quark mass ratio $z$=0.56 can vary in 0.35$\div$0.6 range \cite{PDG08}, the exact interpretation of
experimental results is significantly restricted. We use $S=0.5$ and  $z$=0.56 as reference when calculating the axion flux for
the KSVZ axion model.

According to (\ref{ma}), the constants $g^{0}_{AN}$ and $g^{3}_{AN}$ can be expressed in terms of the axion mass (S = 0.5) as
\begin{eqnarray}\label{gan_numeric}
g^{0}_{AN}&=&-4.03\cdot10^{-8}(m_{A}/\mbox{1~eV}), \\ \label{gan_numeric_2} g^{3}_{AN}&=&-2.75\cdot10^{-8}(m_{A}/\mbox{1~eV}).
\end{eqnarray}

The values of $g_{AN}^{0}$ and $g_{AN}^{3}$ for the DFSZ axion depend on the additional unknown parameter $\cos^2\beta$ which is
defined by the ratio of the Higgs vacuum expectation values, but they have the same order of magnitude \cite{Sre85,Kap85}.

Because the branching ratio $(\omega_{A}/\omega_{\gamma})$ is a model dependent we can consider the parameter $(g_{AN}^0
\beta+g_{AN}^3)^2$ in the expression (\ref{axion_prob}) as a free unknown parameter characterizing the axion-nucleon coupling.

The cross-section of the resonant absorption of the axions is given by the expression similar to the one for the $\gamma$-ray
absorption and corrected by the $\omega_{A}/\omega_{\gamma}$ ratio \cite{Ruj89}:
\begin{equation}\label{crosssection}
\sigma(E_{A})=2\sqrt{\pi}\sigma_{0\gamma}\exp\left[-\frac{4(E_{A}-E_{M})^{2}}{\Gamma^{2}}\right]\left(\frac{\omega_{A}}{\omega_{\gamma}}\right),
\end{equation}
where $\sigma_{0\gamma}$ is the maximum cross-section of the $\gamma$-ray resonant absorption and $\Gamma = 1/\tau$. The experimentally
obtained value of $\sigma_{0\gamma}$ for $^{57}$Fe nucleus is equal to $2.56\times10^{-18}$ cm$^{2}$.

In order to obtain the total cross-section value, we should integrate the expression (\ref{crosssection}) over the axion
spectrum given by (\ref{axionflux_num}). The integration of the narrow Gaussian distribution (\ref{crosssection}) over the wide
axion spectrum yields a value close to $\Phi_{A}(E_{M1})$. Using the dependence of $\Phi_{A}$ (\ref{axionflux_num}) and
$\sigma(E_A)$ (\ref{crosssection})  on the ($\omega_A$/$\omega_{\gamma}$) ratio, and, therefore, on the axion mass $m_{A}$
(\ref{axion_prob},\ref{gan_numeric},\ref{gan_numeric_2}), one can numerically present the estimated rate of resonant absorption
of axions by $^{57}$Fe nucleus ($S$=0.5, $z$=0.56):
\begin{eqnarray}\label{count_speed}
R = 1.56\cdot10^{-3}(\omega_{A}/\omega_{\gamma})^2 \\ \label{count_speed_2} =
5.16\cdot10^{-3}(g_{AN}^0\beta+g_{AN}^3)^4(p_A/p_{\gamma})^6
\\ \label{count_speed_3}  = 9.29\cdot10^{-34}(m_{A})^{4}(p_A/p_{\gamma})^6.
\end{eqnarray}
The amount of the detected $\gamma$-rays following the axion absorption is determined by the target mass, the time of measurement and the
detector efficiency, while the observation probability for the 14.4 keV peak depends on the background level of the experimental setup.

\section{Experimental setup.}
The planar Si(Li) detector with the diameter of the sensitive region 17 mm  and 2.5 mm thick was used for the detection of 14.4
keV $\gamma$-rays. The detector was mounted inside the vacuum cryostat with 20 $\mu$m beryllium window. The 290 mg iron target
enriched to 91\% of $^{57}$Fe isotope was placed directly on the beryllium window, the distance between the detector's surface
and Fe target was $\approx$3 mm. The surface density of the target was 92 mg/cm$^2$, while the attenuation of 14.4 keV
$\gamma$-rays in iron corresponds to 16 mg/cm$^2$.

The detector was surrounded by 12.5 cm copper and 2.5 cm lead shields to eliminate the external $\gamma$-radioactivity. The
background level at 14.4 keV was decreased in 110 times in comparison with the unshielded detector. The detector was located
above the ground surface, so in order to minimize the influence of cosmic radiation and fast neutrons, we used the active
shielding assembled of five $50\times50\times12$ cm organic crystal scintillators. The neutron shielding and active muon veto
cover the top and sides of the passive shield, except for the side where the detector Dewar is located. The anticoincidence veto
signal was obtained from the logical OR of all the photomultiplier tube discriminator outputs. The Si(Li) detector operated in
an anticoincidence mode with plastic scintillators and the background events around 14.4 keV were reduced by a factor of 2.5.
The rate of 50 $\mu$s veto signals was 600 counts/s, that leads to $\approx$3\% dead time. The spectrum of the Si(Li) signals
obtained in the coincidence with veto signals allows checking the probability of excitation 14.4 keV level by the nuclear active
component and cosmic ray muons.

The spectrometric channel of the Si(Li) detector was organized  in the following way. The first-stage field-effect transistor
was mounted on a Teflon block, a few mm from the center contact of the silicon crystal, while the preamplifier was placed beyond
the passive shielding. The data acquisition system was based on standard Camac electronics. A signal produced by the
preamplifier was transmitted to two separate amplifiers with different gain ratios, thus providing the possibility to collect
spectra from both lower (0-60 keV) and higher energy (0-500 keV) regions.   The amplifier outputs were converted using
analog-to-digital converters, controlled by a PC through parallel interfaces. Taking into account the active shielding
coincidence spectra for each amplifier, there were four 4096-channel spectra being recorded.

The energy scale was defined using standard calibration sources of $^{55}$Fe, $^{57}$Co and $^{241}$Am. The energy resolution of
the detector (FWHM)  determined by the 14.4 keV $\gamma$-line of $^{57}$Co source turned out to be 280 eV.

The detection efficiency of the Si(Li) detector was measured with 14.4 keV $\gamma$-rays from a standardized $^{57}$Co source.
The self-absorption of 14.4 keV $\gamma$-rays by the iron target was found via detailed M-C simulation. The overall detection
efficiency for 14.4 keV $\gamma$ is estimated to be (2.30$\pm$0.1)\%.

\section{Results.}
\begin{figure}
\includegraphics[bb = 20 120 500 755, width=8cm,height=9cm]{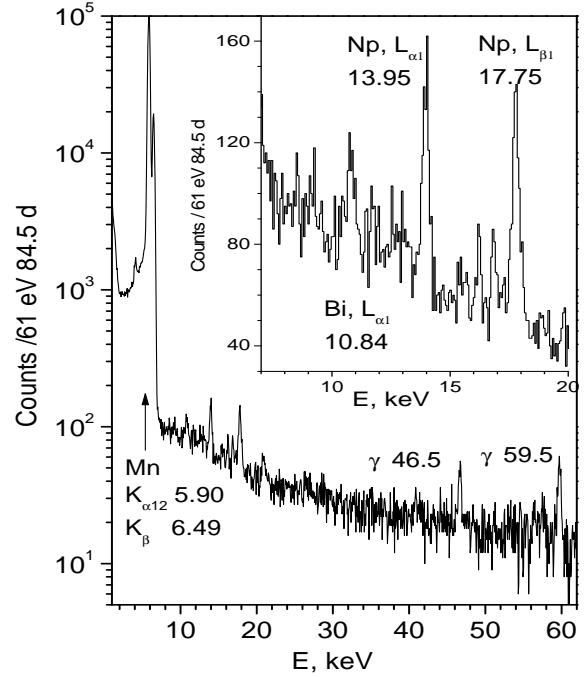}
\caption {Energy spectrum of the Si(Li) detector measured in the region 1-60 keV (main) and 7-20 keV (inset). } \label{fig1}
\end{figure}

The measurements were carried out during 84.5 days of livetime by $\sim$2-hour runs. The obtained energy spectra are given in
Fig.\ref{fig1}. Since the $^{57}$Fe target consisted of a small amount of $^{55}$Fe, the most intense peaks are connected with
the characteristic X-rays of $^{55}$Mn (K$_{\alpha}$= 5.9 keV, K$_{\beta}$=6.49 keV). One can clearly identify several peaks
related to the $^{241}$Am radioactivity. The low-energy region contains the L-series of characteristic X-rays caused by
$^{241}$Am $\rightarrow$ $^{234}$Np$^{*}+\gamma$ decay. The 13.9 keV peak consists of two lines with the energies 13.946 keV
(13\%, L$_{\alpha1}$) and 13.761 (1.4\%, L$_{\alpha2}$), the 17.8 keV peak is more complex one formed by L$_{\beta1-5}$ lines.
The peaks with the energies of 10.84 keV and 13.2 keV (L$_{\alpha1}$ and L$_{\beta1-5}$ of Bi, respectively) are present due to
$^{210}$Pb$\rightarrow^{210}$Bi $\beta$-decay in $^{238}$U series. Intense $^{241}$Am (59.5 keV) and $^{210}$Bi (46.5 keV)
$\gamma$-rays are observed at the higher energy region. All these peaks were used to determine the final energy scale and energy
resolution $\sigma$(E) of the detector.

\begin{figure}
\includegraphics[bb = 20 120 500 755, width=8cm,height=9cm]{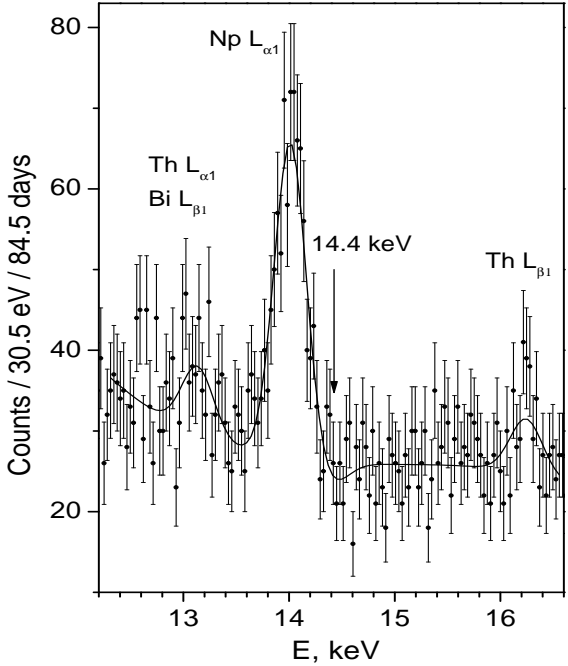} \caption {Fitting results of the energy spectrum inside the 12.4-16.6
keV region. The location of the expected axion peak is denoted by arrow.} \label{fig2}
\end{figure}

Fig.\ref{fig2} shows the detailed energy spectrum within the 12.2 - 16.6 keV interval, where the axion peak was expected.
Apparently, there is no pronounced peak at the 14.4 keV. In order to determine the upper limit for the number of events inside
the expected peak we used the maximum likelihood approach. The likelihood function was determined as a sum of four Gaussians and
the polynomial background. Three Gaussians represent the known characteristic X-rays, one gaussian stands for the expected 14.4
keV axion peak and the third order polynomial is used for the smooth background:
\begin{eqnarray}\label{likelihood}
\nonumber \lefteqn{N(E)=a+b\cdot E+c\cdot E^{2}+d\cdot E^{3}}\\
&&+\frac{1}{\sqrt{2\pi}\sigma}\sum^{4}_{i=1}S_{i}\exp\left[-\frac{(E_{i}-E)^{2}}{2\sigma^{2}}\right].
\end{eqnarray}
Peak positions and energy resolution ($\sigma$) were fixed, while the peak areas and background polynomial coefficients were
independent free parameters. The total number of degrees of freedom at the 12.4 - 16.6 keV region amounted to 250.

The fitting result is given in Fig.\ref{fig2}. The minimum $\chi^{2}$ value corresponds to the nonphysical value of the 14.4 keV
peak area $S=-25$ events. The upper limit for the amount of events inside the peak was found via the conventional approach: the
dependence of $\chi^{2}$ on the peak area $S$ was calculated for various values of $S$ while the rest parameters were left
unrestrained. Then the appearance probability of the given $\chi^{2}(S)$ value was found and the obtained function
$P(\chi^2(S))$ was normalized to unity for the $S\geq0$ region. Thus, the upper limit appeared to be equal to $S_{lim}=24$
events. For the rate of axion absorption $R$ given by (\ref{count_speed}) the expected number of registered 14.4 keV
$\gamma$-quanta is:
\begin{equation}\label{slim}
S=\varepsilon\cdot \eta\cdot N\cdot T\cdot R\leq S_{lim}.
\end{equation}
Here, the number of $^{57}$Fe nuclei $N=2.78\cdot10^{21}$, the measurement time $t=7.30\cdot10^{6}$s, $\gamma$-ray registration efficiency
$\varepsilon=2.3\cdot10^{-2}$ and internal conversion ratio $\eta=0.105$.

The relation (\ref{slim}) limits possible values of the axion-nucleon couplings constants and axion mass. In accordance with
equations (\ref{count_speed}) and on condition that $(p_A/p_{\gamma})^6\cong1$ provided for $m_A < 2$ keV one can obtain:
\begin{equation}\label{limgAN}
|-1.19\cdot g_{AN}^0+g_{AN}^3|\leq 3.12\cdot10^{-6}, \mbox{ and}
\end{equation}
\begin{equation}\label{limma}
m_A \leq 151 \mbox{ eV at 90\% c.l.}
\end{equation}

The limit (\ref{limma}) is the strongest up-to-date result obtained with 14.4 keV solar axions. The previous limit ($m_A\leq$
216 eV \cite{Nam07}) is improved in 1.4 time. Because the expected intensity of 14.4 peak depends on $m_{A}^4$ that in fact
corresponds to an increase the sensitivity of the experiment in 4 times.
\begin{figure}
\includegraphics[bb = 20 120 500 755, width=8cm,height=9cm]{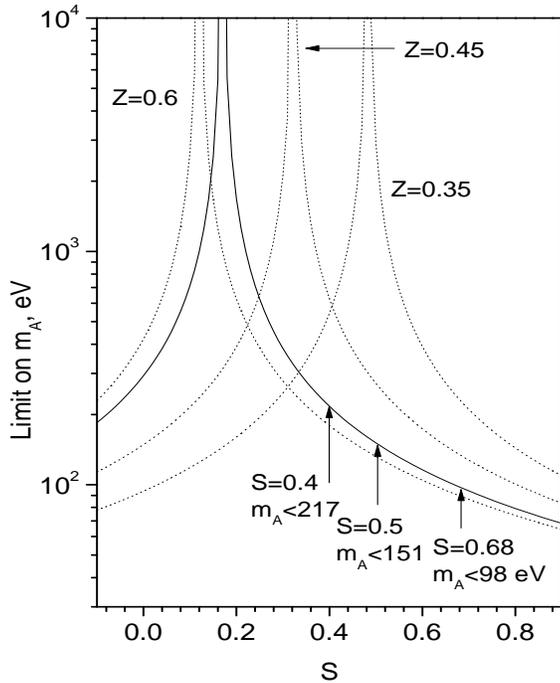}
\caption {The limit on the hadronic axion mass versus $S$ for various  values of $z$. The solid line corresponds to $z$=0.56.}
\label{fig3}
\end{figure}
The obtained limit on hadronic axion mass depends on the exact values of the parameters $S$ and $z$ (Fig.\ref{fig3}). The
uncertainty of the flavor-singlet axial-vector matrix element $S$ changes the obtained constraints significantly: $m_A\leq$217
eV ($S$=0.4) and $m_A\leq$98 eV ($S$=0.68). Moreover, if the value of S is close to 0.17, the limit on the hadronic axion mass
can not be derived from the present experiment.

The value of $u$- and $d$ quark-mass ratio $z$=0.56 is generally accepted for axion papers, but it could vary in the range
0.35$\div$0.6 \cite{PDG08}. For the fixed $S$=0.5, the changes of $z$ from 0.6 to 0.45 lead to the limits
$m_A\leq$(132$\div$272) eV correspondingly. If $z\cong$0.34, the hadronic axion mass can not be restricted by the experiment.
The reason is that the ratio $(\omega_A/\omega_{\gamma})$ vanishes for some values of $z$ and $S$. One can obtain from
(\ref{axion_prob}) and (\ref{couplcons}) that if $S$ and $z$ obey the relation $S=1-1.5\cdot (z\pm 0.01)$, the limit on the
axion mass can not be obtained. Nevertheless, the limit on axion-nucleon coupling given by (\ref{limgAN}) is still valid.

As mentioned above, the main disadvantage of the approach with axions emitted in 14.4 keV M1-transition of $^{57}$Fe is that the
nuclear-structure-dependent parameter $\beta$ has a negative value which, together with a poorly constrained flavor singlet
axial-vector matrix element $S$, leads to large uncertainty of the ($\omega_A / \omega_{\gamma}$) ratio.

\section{Conclusion.}
A search for resonant absorption of the solar axion by $^{57}$Fe nuclei was performed using the planar Si(Li) detector installed
inside the low-background setup.  The intensity of the 14.4 keV peak measured for 84.5 days turned out to be $\leq0.28$
events/day. This allowed us to set the new upper limit on the hadronic axion mass of $m_{A}\leq151$ eV (90\% c.l.) with the
generally accepted values $S$=0.5 and $z$=0.56. The obtained limit strongly depends on the exact values of the parameters $S$
and $z$.

\end{document}